\documentclass[fleqn,twoside]{article}
\usepackage{espcrc2}

\usepackage{epsfig}
\usepackage{graphicx}
\usepackage[figuresright]{rotating}

\newcommand{\AmS}{{\protect\the\textfont2
  A\kern-.1667em\lower.5ex\hbox{M}\kern-.125emS}}

\RequirePackage{xspace}
\def\ccbar {\ensuremath{c\overline c}\xspace}

\def\invnb {\ensuremath{\mbox{\,nb}^{-1}}\xspace}

\def\invpb {\ensuremath{\mbox{\,pb}^{-1}}\xspace}

\def\invfb   {\ensuremath{\mbox{\,fb}^{-1}}\xspace}

\title{Open charm and charmonium production at LHCb}

\author{M. J. Charles\address[Ox]{Department of Physics, University of Oxford, \\
    Denys Wilkinson Building, Keble Road, Oxford OX1 3RH, United Kingdom},
  on behalf of the LHCb Collaboration.}
       
\begin{document}

\begin{abstract}
We see copious production of charmed hadrons in the early LHCb data. 
The status and plans for charm production studies at LHCb are discussed,
and preliminary signals are shown.
\vspace{1pc}
\end{abstract}

\maketitle

\section{INTRODUCTION}

The Large Hadron Collider (LHC) began colliding protons at a centre-of-mass
energy of 7~TeV at the end of March 2010. In the subsequent months, the
luminosity has increased more or less exponentially: 
approximately 16~\invnb had been delivered to each experiment
at the start of the BEACH2010 workshop,
and at the time of writing these proceedings this has risen to
approximately 2~\invpb. Production of charmed hadrons is a natural
topic to probe with these data samples: the cross sections are large
enough to make precise measurements even with limited data samples, and
early studies on charm distributions will improve our understanding
of hadronic events at the LHC for subsequent analyses.
It is anticipated that $\mathcal{O}(70~\invpb)$
will be delivered during the 2010 run, and 
$\mathcal{O}(1~\invfb)$ in 2011~\cite{bib:myers_ichep2010}.
In this note, we discuss the status of measurements of the cross section
for prompt production of open and hidden charm mesons in $pp$ collisions
at $\sqrt s = 7$~TeV in bins of rapidity $y$ and transverse momentum $p_t$.

Charm and beauty hadrons are light compared to the collision energy at
the LHC, so their production is concentrated in the forward region.
This motivates the design of LHCb~\cite{bib:lhcb}, a single-arm forward spectrometer,
which instruments the angular range $15 < \theta_x < 300$~mrad and
$15 < \theta_y < 250$~mrad, corresponding to approximately
$2.0 < \eta < 5.2$, where $\theta_x$ and $\theta_y$ are the $x$
and $y$ projections of the polar angle,  $\eta$ is the pseudorapidity,
and the $z$-axis corresponds to the nominal beam direction.
This coverage is complementary to the other large LHC
experiments~\cite{bib:alice,bib:atlas,bib:cms}.
The LHCb detector includes 
a silicon microstrip Vertex Locator (VELO) which is positioned at
5~mm from the beam during data-taking;
five downstream tracking stations which use a combination of
silicon pixels and drift chambers;
two Ring Imaging Cherenkov (RICH) detectors with three different
radiators to cover the momentum range $2<p<100$~GeV$/c$;
a calorimeter consting of a Scintillator Pad Detector (SPD)
to detect charged particles, an electromagnetic calorimeter
with a preshower detector, and a hadronic calorimeter;
five muon stations;
and a magnet with a field integral of 4~Tm.

Results at the BEACH2010 conference were presented with samples of
2--14~\invnb. Where available, equivalent plots from larger
data samples have been substituted in these proceedings.

\section{MEASUREMENT OF LUMINOSITY}

The integrated luminosity of the data is an essential input to
cross section measurements. The approach used at LHCb is to calibrate
quantities which are expected to be proportional to the instantaneous
luminosity, such as the number of hits in the SPD, with runs in which the
luminosity is measured directly. These calibrations can then be used to determine the
integrated luminosity for a much larger data sample.

Two methods have been used to establish the instantaneous luminosity.
First, the profiles of the beams can be measured directly in the VELO
using beam-gas events, which are recorded continuously by the LHCb
trigger at a low rate for calibration purposes. Given the
bunch-crossing frequency and beam currents, the luminosity can then be
inferred~\cite{bib:lumi}. Second, a dedicated
run can be used for a Van der Meer scan~\cite{bib:vandermeer}.
The uncertainty on the integrated luminosity when calibrated
in this way at LHCb is estimated to be 10\% at present.

\section{CHARMONIA}

\begin{figure}[tb]
\includegraphics*[width=\columnwidth,viewport=5 2 530 340,clip]{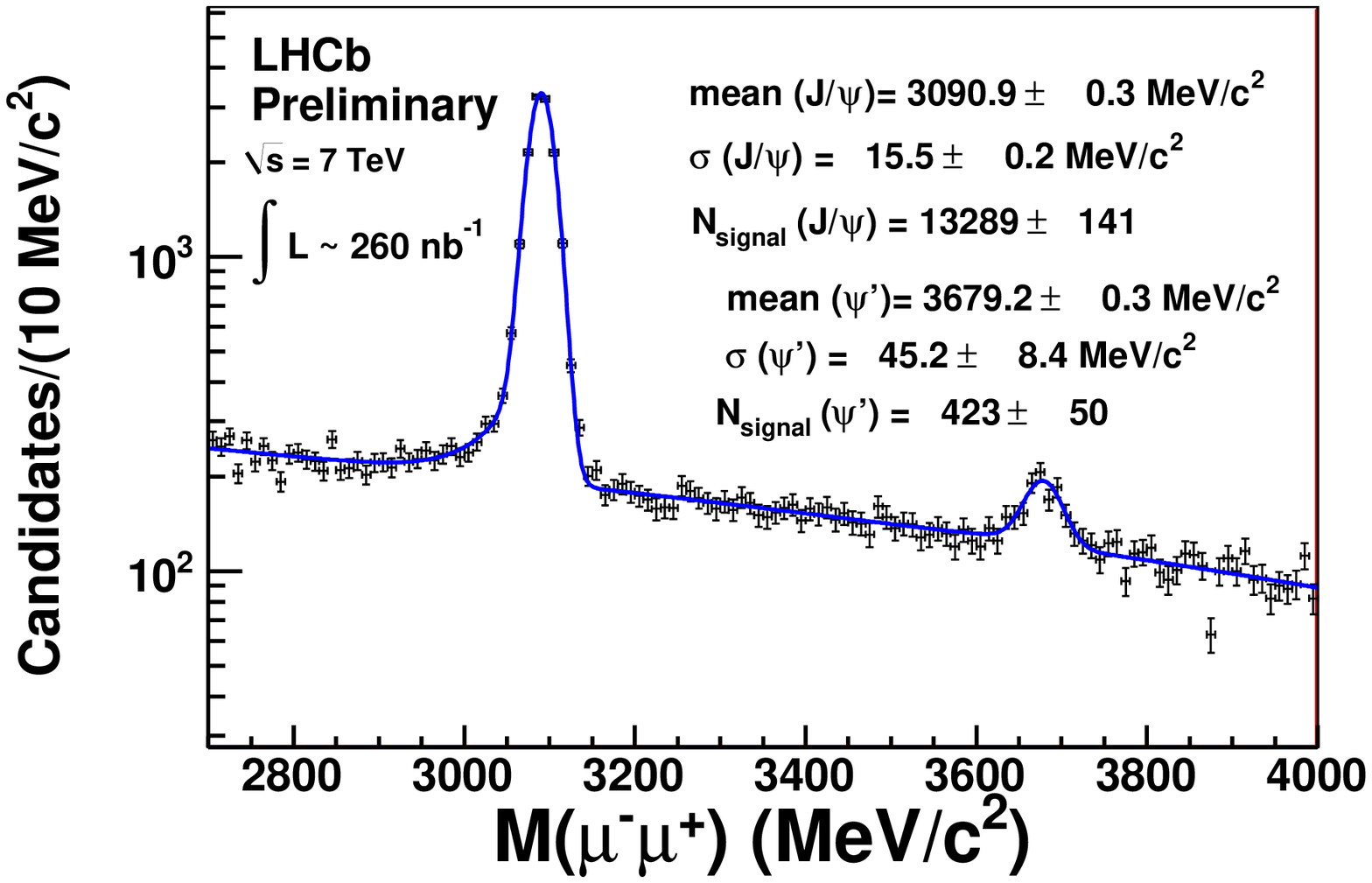}
\includegraphics[width=\columnwidth]{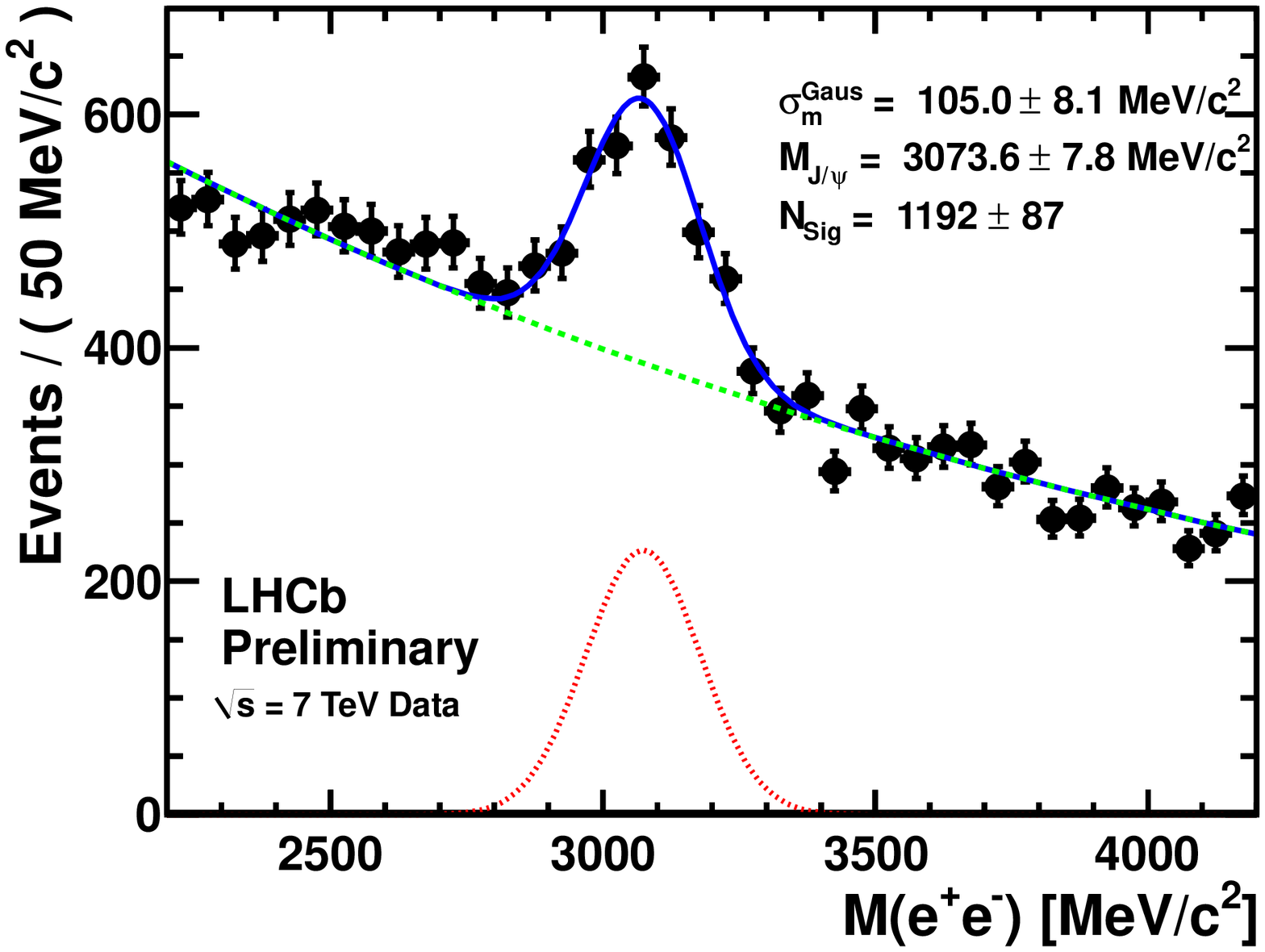}
\caption{
  The invariant mass spectra for $\mu^+\mu^-$ (upper, 260~\invnb)
  and $e^+e^-$ (lower, 150~\invnb).
  The $J/\psi$ and $\psi(2S)$ resonances are visible.
  Note the logarithmic $y$-axis in the upper plot.
}
\label{fig:jpsi_mass}
\end{figure}

\begin{figure}[htb]
\includegraphics*[width=\columnwidth,viewport=31 5 519 261,clip]{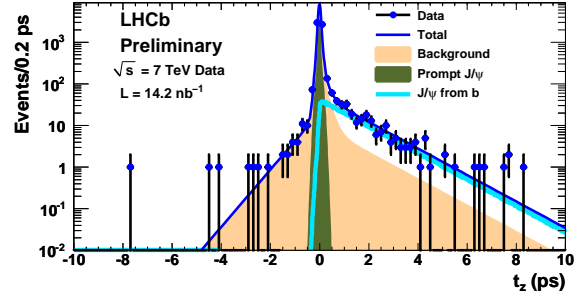}
\caption{
  The pseudo-proper-time distribution for $J\psi \to \mu^+ \mu^-$ candidates
  in 14.2~\invnb of data.
}
\label{fig:jpsi_tz}
\end{figure}

The production mechanisms for \ccbar bound states at hadron colliders
are still far from understood; neither the Colour Singlet Model nor
the Colour Octet Model fully describes the existing experimental data
(see Ref.~\cite{bib:ccReview} and the references therein).
The total cross section, the $p_t$ spectrum, and the polarization of
$J^P=1^-$ \ccbar states are all of interest for probing this problem.

LHCb sees large signals for $J/\psi \to \mu^+ \mu^-$ and $e^+ e^-$, and
a smaller but significant signal for $\psi(2S) \to \mu^+ \mu^-$
(Fig.~\ref{fig:jpsi_mass}). We can divide charmonium
at the LHC into two classes: those produced at an event primary vertex (prompt),
either directly or in the decay of a short-lived resonance such as a
heavier $\ccbar$ state, and those produced in the decay of a long-lived
$B$ hadron (secondary). LHCb plans to measure the cross section as a function
of $y$ and $p_t$ for both classes of production separately.
The two cases can be distinguished using the pseudo-proper-lifetime $t_z$ of
the decay, defined as:
\begin{equation}
  t_z \equiv (d_z M) / p_z ,
\end{equation}
where 
  $d_z$ is the projection in the $z$ direction of the distance between
    the decay vertex and the associated event primary vertex,
  $M$ is the reconstructed invariant mass, and
  $p_z$ is the $z$ component of the momentum.
For prompt production, $t_z$ is described by a narrow peak at zero
with width equal to the resolution; for secondary production, it is
described by an exponential with effective lifetime approximately
equal to the average $b$-hadron proper lifetime. This is illustrated
for $J/\psi$ in Fig.\ref{fig:jpsi_tz}. The fit also includes a 
component for combinatoric background, determined from a mass
sideband above the $J/\psi$. The data clearly require a secondary
component to be present.


\begin{figure}[tb]
\begin{center}
\begin{tabular}{c}
  \includegraphics*[width=0.8\columnwidth]{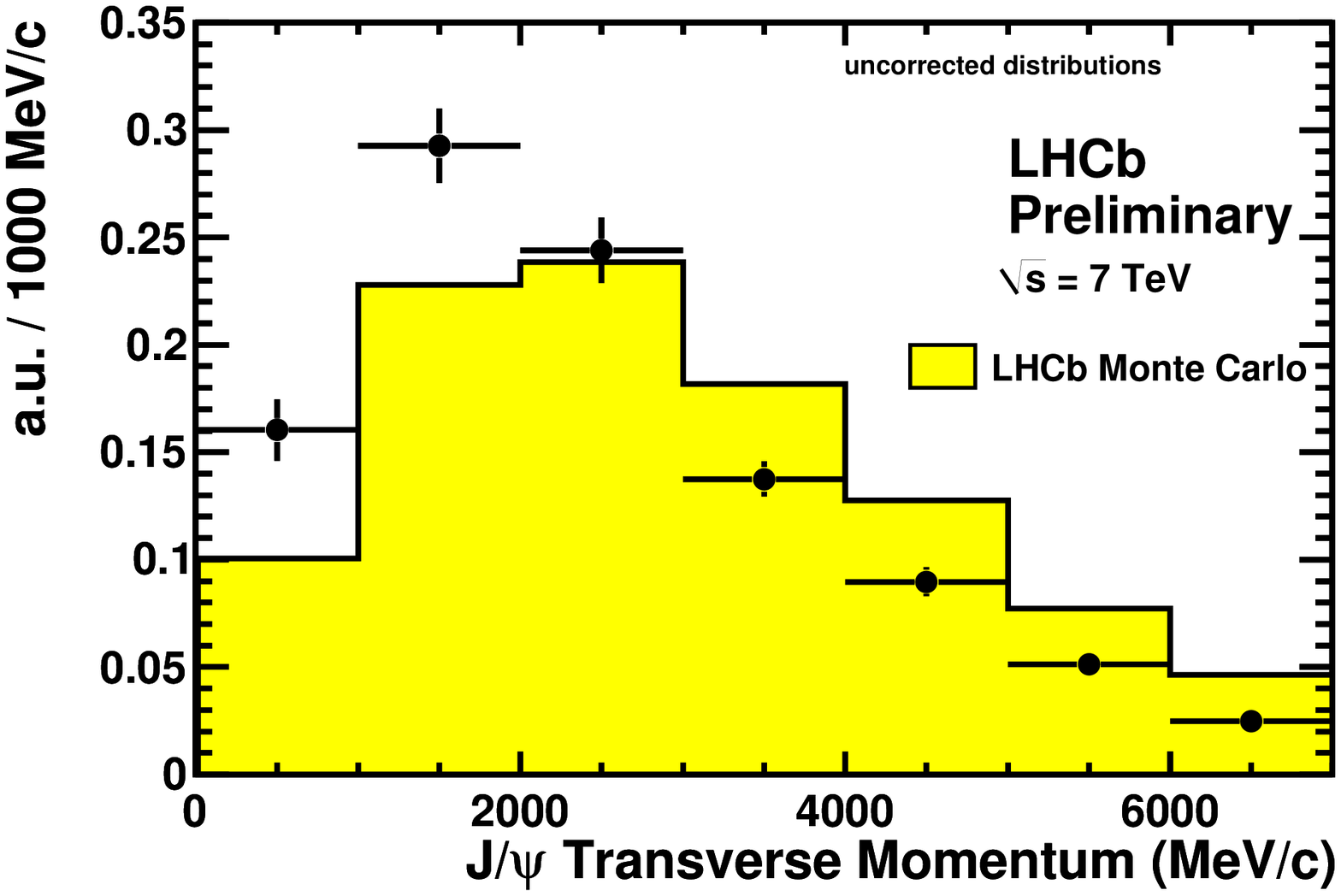} \\
  \includegraphics*[width=0.8\columnwidth]{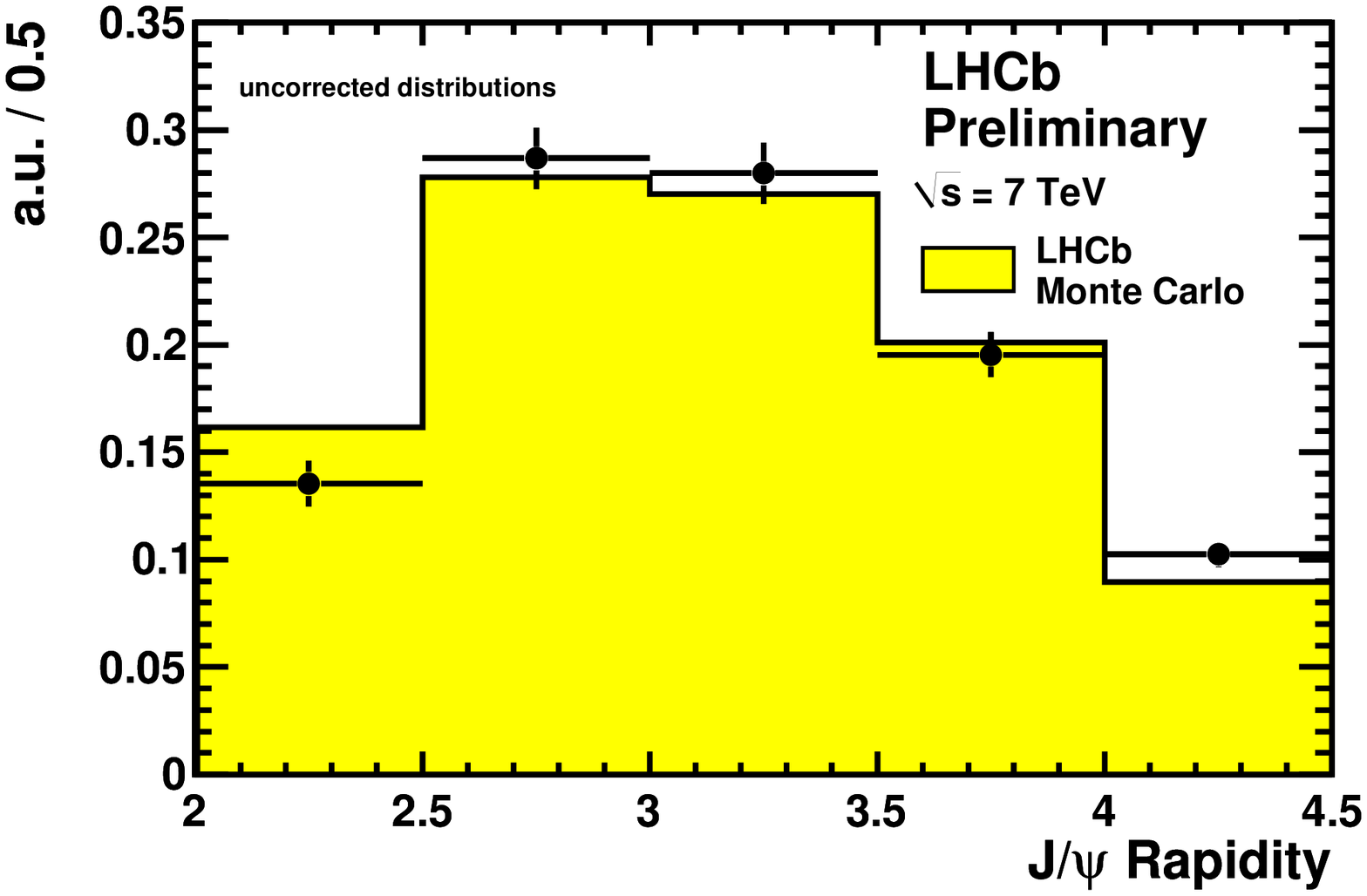}
\end{tabular}
\end{center}
\caption{
  $J/\psi$ yields from 14~\invnb of data, not corrected for efficiency
  and including both prompt and secondary components. 
  The upper plot shows the yield in bins of $p_t$ for $2.5<y<4.0$, and
  the lower plot shows the yield in bins of $y$ for $0<p_t<12$~GeV$/c$.
  The points with error bars show the measured yield with statistical
  uncertainties only, and for comparison the shaded histogram shows
  the distribution seen in the LHCb Monte Carlo.
}
\label{fig:jpsi_slices}
\end{figure}

A measurement in fine-grained bins (0.5 in $y$ $\times$ $1$~GeV$/c$ in $p_t$)
will require $\mathcal{O}($10--20~\invpb), but preliminary studies in
one-dimensional bins are underway: the uncorrected yields
seen in one-dimensional bins are shown in Fig.~\ref{fig:jpsi_slices}.
Firm conclusions cannot be drawn without systematic uncertainties, but these
initial results illustrate the potential improvements from retuning
the Monte Carlo generator with early data.

\begin{figure}[!htb]
\begin{center}
\includegraphics*[width=0.8\columnwidth]{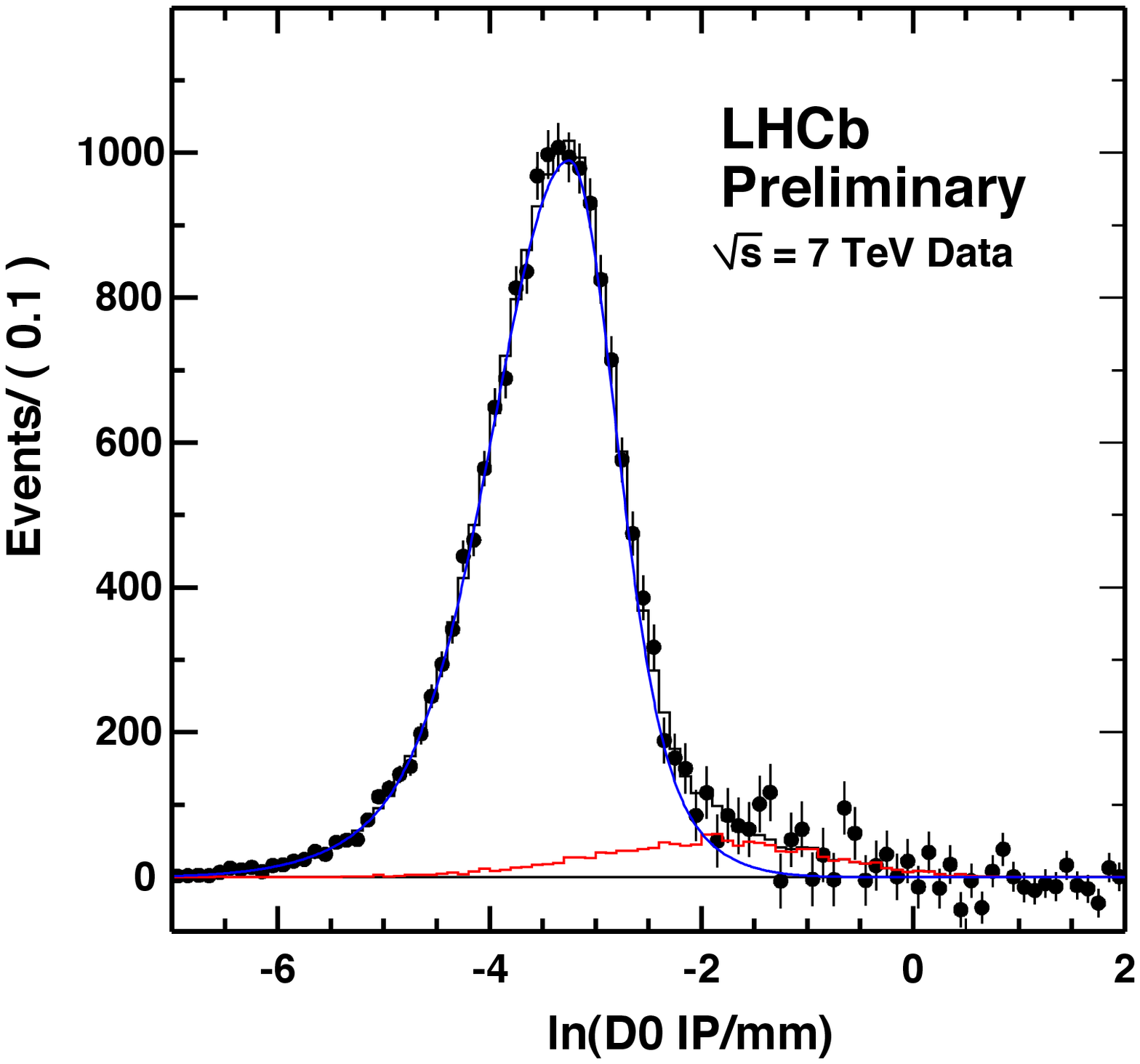}
\includegraphics*[width=0.8\columnwidth,viewport=0 0 286 200,clip]{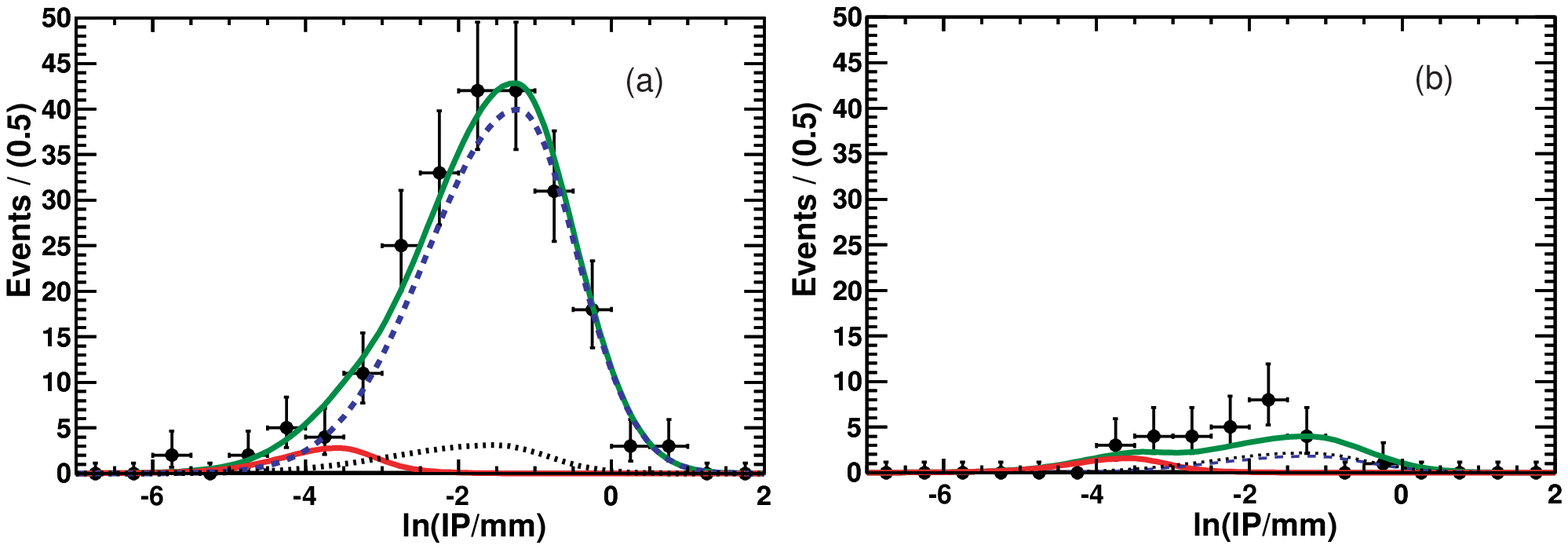}
\end{center}
\caption{
  The logarithm of the $D^0$ impact parameter distribution for
  $B \to D^0 (K^- \pi^+) \mu^{\pm}$ candidates.
  The upper plot (3~\invnb) shows the background-subtracted
  distribution for inclusive $D^0$ candidates; 
  the histograms indicate the fitted prompt and secondary components.
  The lower plot (11~\invnb) shows $D^0$ candidates after requiring
  a $B \to D^0 \mu^+$ tag, which suppresses the prompt
  component strongly;
  the curves indicate the components of the fit:
  prompt (solid), secondary (dashed), background (dotted),
  and their sum (solid).
}
\label{fig:charmLogIP}
\end{figure}

\section{CHARMED MESONS}

\begin{figure*}[htb]
\begin{center}
\epsfig{file=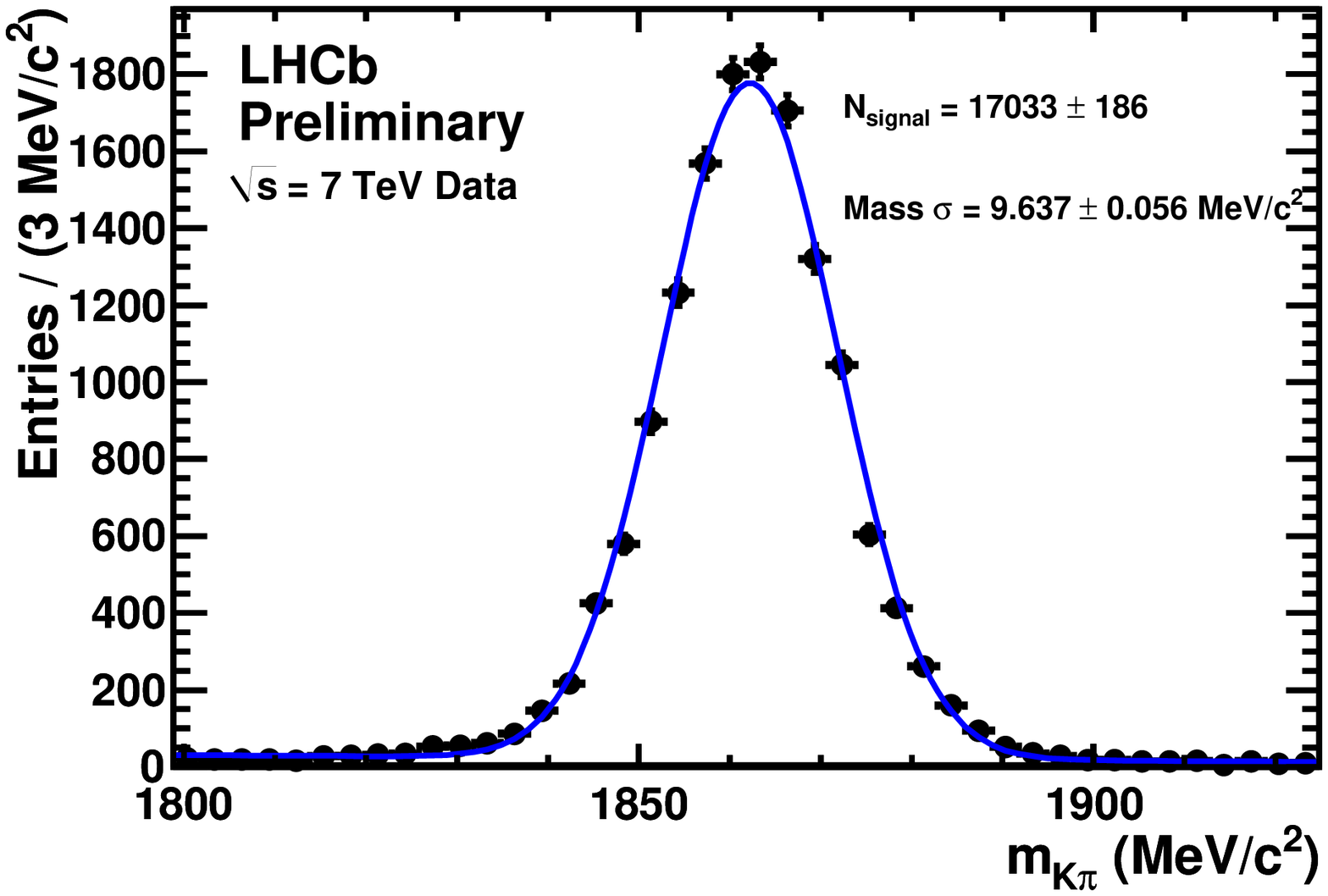,width=0.32\textwidth}
\epsfig{file=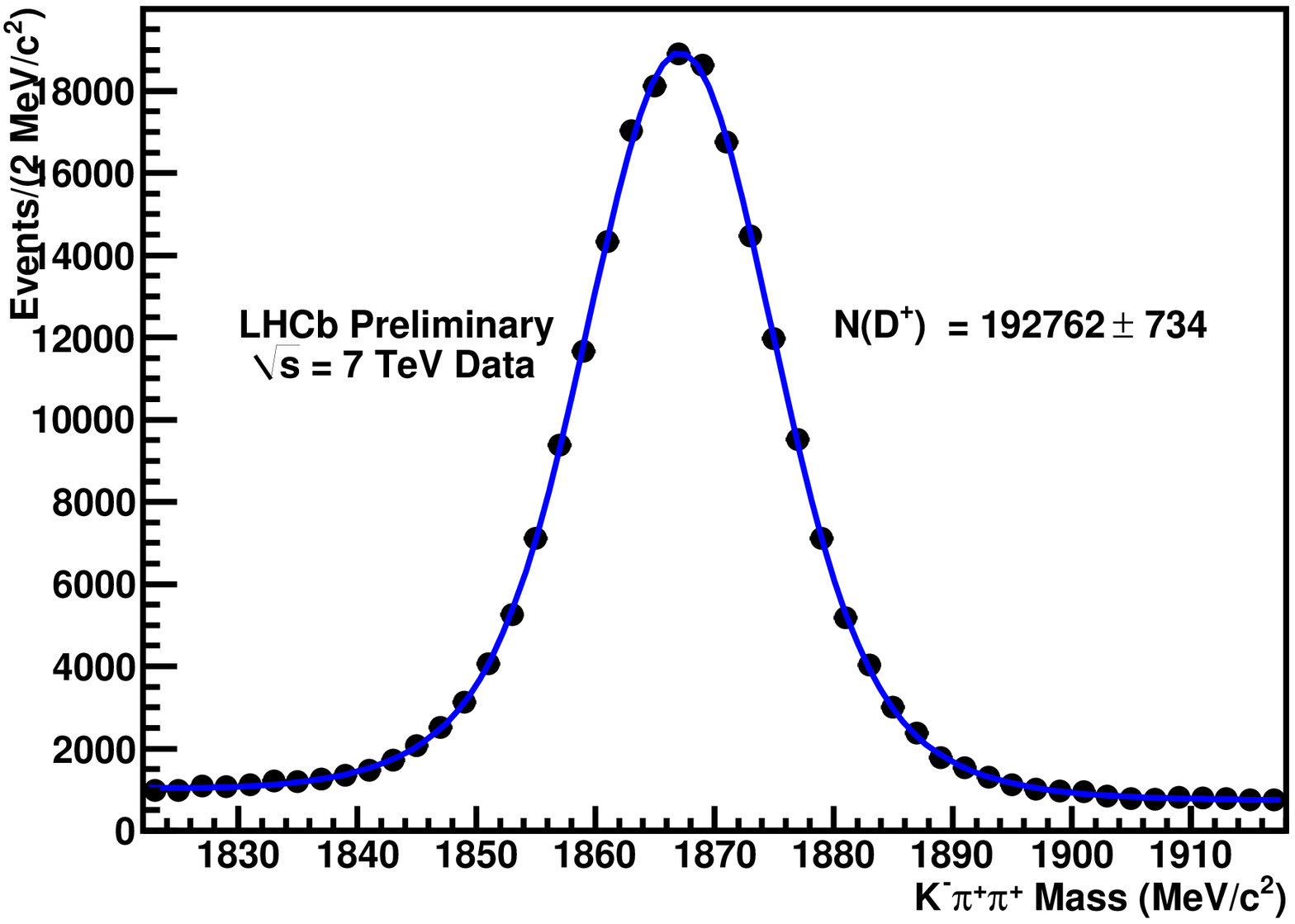,width=0.32\textwidth}
\epsfig{file=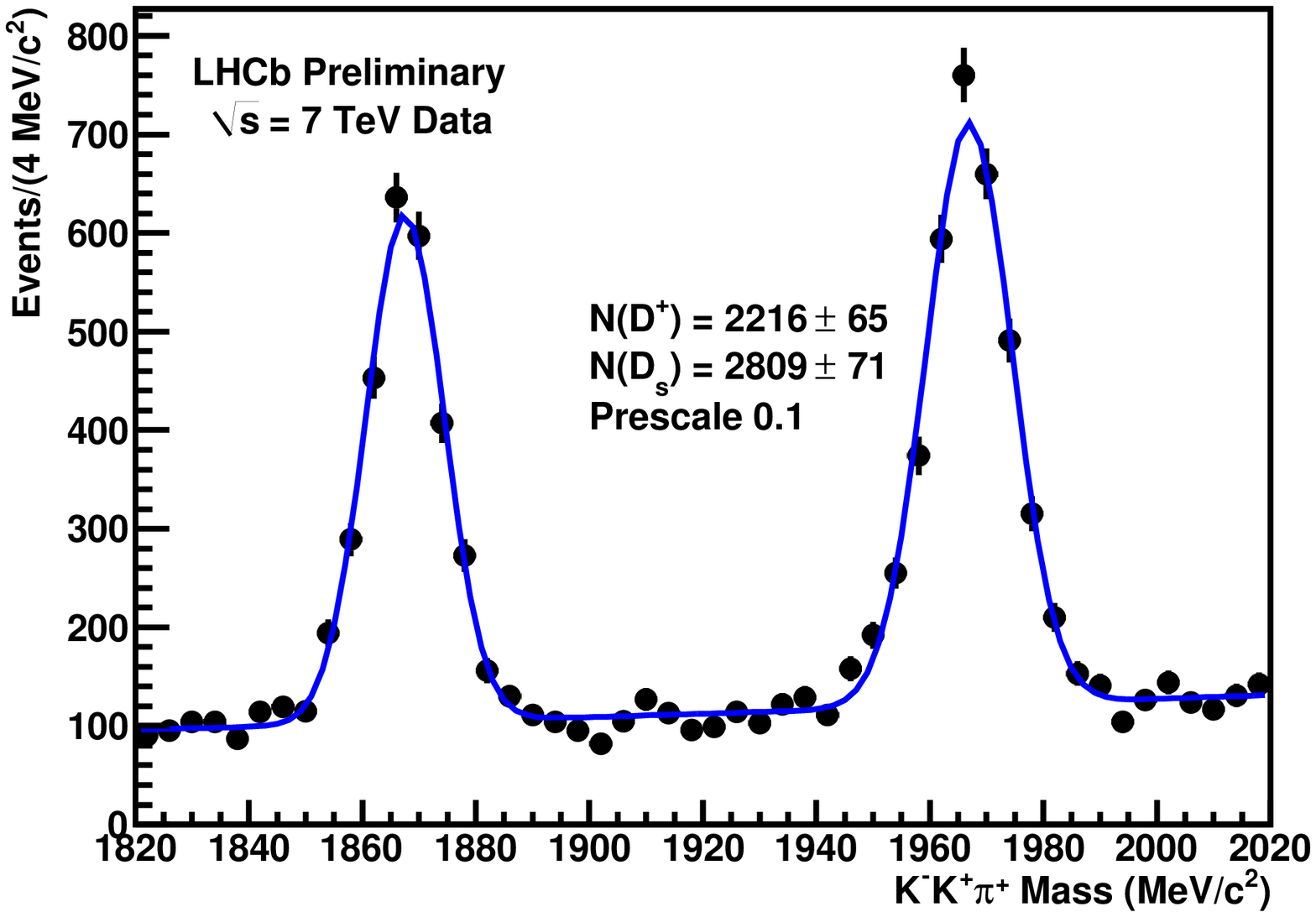,width=0.32\textwidth}
\end{center}
\caption{
  Invariant mass spectra for charmed mesons. From left to right:
  $m(K^- \pi^+)$ for $D^{*+} \to D^0 (K^- \pi^+) \pi^+$;
  $m(K^- \pi^+ \pi^+)$;
  $m(K^- K^+ \pi^+)$.
  The first two plots use 125~\invnb, and the third uses 13~\invnb.
}
\label{fig:charm_mass}
\end{figure*}

Similar production studies are in progress for the charmed mesons
$D^0$, $D^+$, $D^{*+}$, and $D_s^+$. The excellent
signal quality is illustrated in Fig.~\ref{fig:charm_mass}.
As with the $J/\psi$, both prompt and secondary production are
expected at LHCb. However, because the $D^0$, $D^+$, and
$D_s^+$ fly a significant distance at LHCb, separating the two
sources is more complicated. The discriminating variable used is the
impact parameter (IP) of the $D$ candidate with respect to the
associated event primary vertex---prompt charm points back to
the primary vertex and so its IP distribution is dominated by
the resolution, whereas secondary charm in general has a much
larger IP. This is illustrated in Fig~\ref{fig:charmLogIP}.
By fitting both the invariant mass spectrum and the
$\log(\mathrm{IP})$ distribution, the production rate of prompt
charm mesons can be measured. Inclusive cross section measurements
in bins of $p_y$ and $y$ using this technique are now underway.


\section{OUTLOOK}

The production measurements described above are in progress and will be
finalized over the coming months. In the meantime, the LHC continues to
deliver ever-increasing luminosities. With the much larger data samples
expected by the end of the 2010 run, LHCb will be able to search for new
physics in charm decays. Three key analyses will be 
measurements of $y_{CP}$ and $A_{\Gamma}$ with $D^0 \to K^- K^+$ and~$K^- \pi^+$; 
searches for direct CP violation in $D^+ \to K^- K^+ \pi^+$ and $\pi^- \pi^+ \pi^+$; and
a search for $D^0 \to \mu^+ \mu^-$. 
These studies will benefit greatly from the combination of copious charm
production, a large boost (especially valuable for $y_{CP}$ and $A_{\Gamma}$),
and excellent charged particle identification across the full momentum range.

\end{document}